\documentstyle[12pt,aaspp4]{article}

\def\gtsim{\ {\raise-0.5ex\hbox{$\buildrel>\over\sim$}}\ }
\def\ltsim{\ {\raise-0.5ex\hbox{$\buildrel<\over\sim$}}\ }

\begin{document}

\title{Optical Counterparts of X-Ray Point Sources Observed by CHANDRA in
NGC5128: 20 New Globular Cluster X-Ray Sources
\footnote{Based on observations collected with the 
Very Large Telescope of the European Southern Observatory,
with the Magellan I Baade Telescope
of the Carnegie Institution, and with the Cerro Tololo Inter-American Observatory
0.9 m telescope, on images of the Very Large Telescope obtained 
from the ESO Archive, and on images obtained from the NASA/ESA Hubble Space 
Telescope Archive.
}}

\author{
 Dante Minniti $^{1}$,
 Marina Rejkuba $^{2}$,
 Jos\'e G. Funes, S.J.$^{3}$,
 Sanae Akiyama$^{4}$
}

\altaffiltext{1}{Department of Astronomy, P. Universidad Cat\'olica, 
Av. Vicu\~na Mackenna 4860, Casilla 306, Santiago 22, Chile\\
E-mail:  dante@astro.puc.cl}

\altaffiltext{2}{European Southern Observatory, Karl-Schwarzschild-Str. 2, D-85748 Garching b.  M\"{u}nchen, Germany\\
E-mail: mrejkuba@eso.org}

\altaffiltext{3}{Vatican Observatory Research Group, Steward Observatory, University of Arizona, Tucson AZ 85721, USA\\
E-mail: jfunes@as.arizona.edu}

\altaffiltext{3}{Steward Observatory, University of Arizona, Tucson AZ 85721, USA\\
E-mail: sakiyama@as.arizona.edu}

\begin{abstract}
VLT images in $BVI$ are used to identify the optical counterparts to
bright CHANDRA X-ray points sources discovered by Kraft et al.
(2001, ApJ, 560, 675) in NGC5128.  Of a total of 111 X-ray point sources 
with $L_X>2\times 10^{36}$ ergs s$^{-1}$ present in a 56 arcmin$^2$ 
field centered on this galaxy, 58 have optical 
counterparts. Based on the sizes, optical magnitudes and colors, 20 new globular
cluster counterparts of X-ray sources are identified, plus 3 identified based 
on their sizes. This brings the total number of globular cluster X-ray sources
in this galaxy to 33, and establishes that $30\%$ of the X-ray point sources
in NGC5128 are associated with globular clusters.  
These X-ray globular clusters occupy the brightest end of the
globular cluster luminosity function, indicating that 
bright low-mass X-ray binaries are preferentially found in massive
clusters. Most of the globular clusters with 
X-ray sources have red colors, with $1.0<V-I<1.5$, indicating that
low-mass X-ray binaries are preferentially formed in metal-rich clusters.  
The NGC5128 X-ray globular cluster sources are brighter in comparison
with the Milky Way sources:  there are 24 globular clusters with X-ray 
sources of $L_x>10^{37}$ erg sec$^{-1}$.   There is, however, no
globular cluster X-ray source in NGC5128 as bright as expected for
an accreting black hole.  
In addition, 31 optical counterparts of X-ray point sources
that are not associated
with globular clusters are identified. Finally, 53 X-ray point
sources (48\% of the population), do not have any optical counterparts 
down to the faintest magnitude limits ($B=25$).
\end{abstract}


\section{Introduction}

The peculiar giant elliptical galaxy NGC5128 (Centaurus A)
is bigger than the Moon on
the sky at optical wavelengths (measuring $1\times 2$ deg$^2$), in radio
(with outer lobes separated by $8$ deg), and in X-rays (with diffuse
emission extending $1/2$ deg). This galaxy
holds the largest globular cluster system
within several Mpc, providing a good basis to understand more distant giant
galaxies at all wavelengths.

The bright X-ray point sources discovered by CHANDRA in distant 
elliptical galaxies are mostly low-mass X-ray binaries (LMXBs), and some
are associated with globular clusters (Kraft et al. 2001, Sarazin et al. 2001,
Angelini et al. 2001, Kundu et al. 2002).
In fact, globular clusters could be the birthplaces for all LMXBs, as
recently proposed by White et al. (2002).
A complete census and optical identification of the population of globular
cluster X-ray sources in distant ellipticals is not possible due
to their large distances: only the very brightest sources are detected. 
NGC5128 is the nearest example of a giant
elliptical galaxy that contains a populous globular cluster
system, a prime opportunity for the identification of optical
counterparts to CHANDRA X-ray sources. No other giant elliptical
galaxy is close enough to identify all X-ray sources  
down to luminosities of $L_X\approx 10^{36}$ ergs s$^{-1}$
(Kraft et al. 2001), while at the same time reaching the 
faint end of the globular cluster luminosity function (Rejkuba 2001),
in order to make an adequate census of the globular cluster X-ray sources.

In the first comprehensive X-ray study of the giant elliptical galaxy NGC5128,
Kraft et al. (2000, 2001) detected 246 bright X-ray point sources with CHANDRA,
using two overlapping observations
with the ACIS-I which has a 16'x 16' field of view
These X-ray point sources with $L_X> 10^{36}$ ergs s$^{-1}$ in
the band [1-3 keV] 
are supposed to be either X-ray binaries or SN remnants.
They searched for optical counterparts using the Digital Sky Survey, and
identified 8 of the 246 point sources
 as background stars, and other 9 that matched
the positions of known globular clusters from the lists of Harris et al.
(1992) and Minniti et al. (1996).

In our galaxy, a dozen X-ray sources have been detected in globular
clusters (Verbunt et al. 1995).  All of them are called bright X-ray sources, with
luminosities $L_X> 10^{36}$ ergs s$^{-1}$ as defined by Verbunt et al.
(1995).  These bright X-ray sources are thought to be accreting neutron stars,
and so far no evidence for an accreting black hole has been found among the
Milky Way globular cluster X-ray sources.  
Only one source --
NGC6624 -- has $L_X> 10^{37}$ ergs s$^{-1}$ (Verbunt et al. 1995).

So far the galaxy with more globular cluster X-ray sources known is M31.
Di Stefano et al. (2002) identified 28 globular cluster X-ray sources in M31 using 
CHANDRA. 
They found also that the globular cluster X-ray source population in 
M31 is brighter in general than that of the Milky Way.
Clearly, the Milky Way sample is small to allow statistical studies, 
mainly due to the presence of only 150 globular clusters in our galaxy. 
A larger galaxy like NGC5128, hosting 1700 globulars (Kissler-Patig 1997), 
allows a 10$\times$ improvement
in the study of the X-ray source population in globular clusters.

In this paper we report the identification of optical counterparts
to the bright X-ray point sources in the inner regions of NGC5128
based on VLT optical images.  These images cover the central 
$7.7\times 7.7$ kpc
of this galaxy (Figure 1), well within one effective 
radius $R_e=5.24$ kpc (Dufour et al. 1979). Based on their colors and 
magnitudes, a large fraction of the sources are associated with globular
clusters. We discuss the properties of these sources and compare them with
similar sources in the Milky Way and other galaxies.

\section{Observations and Reductions}

The images of the inner regions of
NGC5128 illustrated in Figure 1 were taken from the European Southern Observatory
(ESO) Archives.
These 300 sec image in the $B$-band, 240 sec image in the $V$-band,
and 300 sec image in the $I$-band filters 
covering the central $6\farcm8 \times 6\farcm8$ of NGC5128,
were acquired with FORS2 at the ESO Very Large Telescope (VLT) UT4 Kueyen.
The ESO pipeline reduction of the frames was adopted, although we note that
the images were taken far appart from the calibrations.
Thus, the zeropoints for our photometry are taken from a comparison with 
the globular cluster photometry of Harris et al. (1992, 14 objects in common),
Tonry \& Schechter (1990, 10 objects in common), and Holland et al. (1999,
17 objects in common). The photometry of Harris et al. (1992) and
Tonry \& Schechter (1990) agree well with each other, as discussed by the latter
authors, and we adopt their calibrations. We caution, however, that these
zeropoints differ from those of 
Holland et al. (1999) by $0.35$ mag in $V$ and $0.45$ mag in $I$, respectively,
possibly due to the different aperture corrections adopted.
The photometry was performed following
standard procedures in IRAF using DAOPHOT with non-variable point spread function. 
The photometry of the field yields 2400 sources with good quality ($chi<2$, and
$-1>sharp<1$) in all filters.  The magnitude limits for the $B$, $V$, and $I$ 
filters are $25$, $24$, and $22$, respectively, except for the inner 
crowded regions with very high surface brightness, where the limiting 
magnitudes can be $1$ magnitude brighter.  Some of the sources 
-in particular the globular clusters- are
resolved, and we have adopted a mean aperture correction for all sources.
The pixel scale of the images is $0.2$ arcsec pix$^{-1}$. The frames were 
registered to the position of the $B$-band image, and the centroids of
the sources in the different filters were matched to within one pixel
in order to obtain colors. 

A map of all the sources detected in the $B$-band image (regardless of 
their color) of the 
central field is shown in Figure 2. Our field corresponds to 
$7.7 \times 7.7$ kpc$^2$ projected at the distance of NGC5128 (D=3.9 Mpc). 
The sizes of the points are proportional to the $B$-band
magnitudes, although saturated objects with $B<19$ are not plotted. 
There is a concentration of optical sources in the
inner dust lane of this galaxy, showing an apparent disky distribution.
The positions of the CHANDRA point sources from Kraft et al. (2001) 
discussed in the following sections are indicated with crossed squares.
The distribution of optical point sources is markedly different
from that of the X-ray point sources.

The $V$ $vs$ $B-V$ and the $I$ $vs$ $V-I$
color-magnitude diagrams for this central field
are shown in Figure 3. The apparent wide color distributions seen in
Figure 3 are real, and mostly due to a combination
of intrinsic color differences of the sources and differential
reddening.  The CHANDRA point sources from Kraft et al. (2001) 
discussed in the following sections are plotted with large black circles.

In addition, we have used five 30 sec exposures of a field centered 3 arcmin
N of the center, obtained without filters with the VLT FORS1 instrument, as 
pre-imaging for one of our spectroscopic runs. 
There is a 80 \% overlap between these and the central field, yielding 
a total area of 56 arcmin$^2$.
While these images are not
calibrated and do not provide photometry, they allow us to identify optical 
sources. In some cases, these optical sources are barely extended, and most
likely are globular clusters of NGC5128.

\section{Identification of Globular Clusters}

NGC5128 is close enough to allow the identification of globular clusters
based on their sizes: at $D=3.9$ Mpc the typical clusters should be
resolved from the ground under good seeing conditions ($<1$ arcsec).
Minniti et al. (1996) and Alonso \& Minniti (1997) identified globular clusters
in NGC5128 based on their resolved sizes on infrared observations
with the ESO La Silla 2.2m telescope.
Holland et al. (1999) also identified NGC5128 globular clusters based on their
sizes using the Hubble Space Telescope.
Rejkuba (2001) used the ESO VLT to find clusters in
this galaxy based on their sizes in the $U$ and $V$-bands, discovering
for the first time clusters
down to the faintest end of the luminosity function.

We measured the sizes of sources in the VLT images, in order to discriminate
globular clusters from foreground stars and background galaxies 
(see Minniti et al. 1996, Alonso \& Minniti 1997). 
The sizes of the globular clusters are $2.5<FWHM_{GC}<3.5$ pixels, with a mean
of $FWHM_{GC}=3.1$ pixels, or $0.62$ arcsec in the $B$-band image. Instead, 
the stellar sources have a mean size of $FWHM_*=2.5$ pixels, or $0.5$ arcsec,
with a very small spread ($2.4<FWHM_{GC}<2.6$ pixels).
Accounting for the size of point sources, the mean measured 
size of NGC5128 globular clusters in the VLT images corresponds to 6.9~pc,
in agreement with the sizes of typical of Milky Way globular clusters. 
A caveat is that these sizes can be adequately measured for clusters with 
$B<23$, for fainter sources the size discrimination is not reliable. 



The color-magnitude diagrams of Figure 3 show
numerous (about 200) sources with
color and magnitudes consistent with globular clusters members of NGC5128
(adopting $E(B-V)$ from Rejkuba 2001).
For a distance of $D=3.9$ Mpc, the brightest NGC5128 globular clusters
with $M_V=-11$ will reach $V=17$, while the faintest ones with $M_V=-3$
will have $V=25$, which is one magnitude fainter than
our limit. The VLT images will miss 
the brightest globulars because saturation sets at about $V=18$.
Indeed, there are 5 saturated sources in the VLT
images that show X-ray emission, but these bright candidates in the 
field have been identified previously as globular clusters
(Sharples 1988, Harris et al. 1992, Minniti et al. 1996). 
The mean population is well sampled, as the peak of the globular cluster
luminosity function at $M_V=-7.4$ will be seen with $V=20.9$.
Their colors should match the 
colors of Galactic and M31 globular clusters, as defined by Rejkuba (2001).
We select as globular clusters all VLT sources matching the following
photometric criteria: $18<V<24$, or $17.5<I<22.5$, and
$0.7<B-V<1.5$, or $0.8<V-I<1.8$. We do not allow for much more reddened
clusters, because these are located preferentially in the inner, crowded 
regions, where source identification is complicated.

The $V-B$ $vs$ $V-I$ color-color diagram of this field outside
the dust lane is shown in Figure 4, with the CHANDRA point sources
again shown as large black circles.
The majority of these sources with well measured $B-V$ and $V-I$ colors
lie in the region occupied by globular clusters.
Sources 36, 77, and 63 have $V-I$ colors typical of globular
clusters, but are too faint and too blue in $B-V$. Likewise, source
90 has a globular cluster $V-I$-like color, but it is too faint and
too red in $B-V$. Therefore, these sources are eliminated from the
list of candidate globular clusters.

Ultimately, the final confirmation of a globular cluster identity should be
spectroscopic.  As part of a different program (see Minniti \& Rejkuba 2002),
spectroscopic observations of three dozen of these
inner NGC5128 globular clusters were taken with the
Magellan I Baade 6.5 m telescope at Las Campanas Observatory
on the nights of May 7 and 8, 2002, 
and with the ESO VLT in service mode on the nights of July 7 and 8, 2002.
Our selection of targets for the spectroscopy was based on colors, but no
objects with $V-I=2$ were observed.

The Magellan I spectra of these objects were taken 
with the Boller \& Chivens spectrograph.
The measured seeing varied between 0.5 and 1.5 arcsec,
and a slit width of 1 arcsec was used.  
We obtained two exposures of 1200 seconds each on source
in order to eliminate cosmic rays. 

The VLT spectra were acquired with the FORS1 spectrograph in
MOS mode, 
under 1 arcsec seeing,
and with a slit width of 1 arcsec. Two exposures of 1800 sec each with
grating G300B and two exposures of 1800 sec each with grating G600V were
obtained.

The spectral reductions and measurements were carried out in IRAF,
using the set of packages in CCDRED and TWODSPEC.
Inspection of the spectra reveals that all these suspected globular
clusters have indeed typical globular cluster spectra.

Unfortunately, only 6 of these globulars for which we obtained spectra
show X-ray emission. In addition, there are 5 other clusters that
have radial velocities measured by Sharples (1988, see Harris et
al. 1992), bringing the total number to 11 X-ray clusters that have
spectroscopic confirmation.  Even though there are spectra
of only a few clusters (1/3 of the sample), the spectra serve as an extra check
to confirm that the identification of fainter globular
clusters based on the sizes, magnitudes and colors is realistic.

\section{Identification of Optical Counterparts to CHANDRA Sources}

Table 1 lists all the CHANDRA point sources present in 
the inner 
$6\farcm8 \times 6\farcm8$
field studied here, in total 97 sources.
These include two point sources in the jet from Kraft et al. (2002):
their sources CX1 and EX1.
Column 1 gives our identification number, followed by RA and DEC (J2000)
as listed by Kraft et al. (2001). 
We caution that these positions are accurate
only to 1-2 arcsecs -- see their paper for details. 
Columns 4 and 5 give the 
X-ray luminosities measured by Kraft et al. (2001). 
$L_X316$ refers to the CHANDRA observations from December 05 1999 and 
$L_X962$ that from May 17 2000.
They adopted a distance $D=3.5$ Mpc, while we prefer $D=3.9$ Mpc
(Rejkuba 2003). 
Therefore, the luminosities listed in Table 1
have been corrected to $D=3.9$ Mpc. Columns 6 and 7 list the
x-y positions in pixels in the VLT images.
The sources that have been identified in the VLT images have magnitudes
and photometric errors listed in columns 8 to 13 of Table 1, and are marked
in Figure 2 with crossed squares. About 1/2 of the X-ray sources do not 
have optical counterparts, but are listed in Table 1 for completeness. 
Finally, the last column indicates foreground stars,
saturated clusters, or H$_{\alpha}$ emission line objects.

In order to find the matches to the coordinates listed by Kraft et al. (2001),
we apply a small shift 
(1 arcsec East and 2 arcsec South) to the VLT images in order to match the
CHANDRA positions.
After this zero point shift, we find all the sources with positions matching 
within 4 pixels, or $0.8$ arcsec.  We achieve a relative astrometric 
accuracy of $0.3$ arcsec rms.
There are 50 matching sources in the VLT images. The number of matches is not
very sensitive to the matching radius.  Four of the sources have two optical 
sources within the $0.8$ arcsec matching radius. 
In these cases, we always adopt the brighter one as the correct match.
While picking the brighter source is a clear bias, it does not affect
our results.

Figure 2 shows a map of all the VLT optical point sources in the central
field, with the
positions of the CHANDRA point sources shown as crossed squares. 
These sources are mostly three types: NGC5128 globular clusters 
located in a spherical
distribution, NGC5128 young clusters (along with SN remnants and WR stars) 
located along the dust disk, and foreground stars of our own galaxy
that should be uniformly distributed. 
The latter are numerous, due to the low latitude of this galaxy.

The density of optical sources in the inner regions of NGC5128 is high, and
the positions of the CHANDRA and VLT images are uncertain enough that
some spurious matches could be expected. We have done a few simulations to see 
how many matches are expected at random given the density of X-ray and optical
point sources and given the size of our search box of 2 arcsec$^2$. These yield
only 3-5 false matches, showing that the optical counterparts of X-ray point
sources listed in Table 1 are real.

Table 2 lists 16 additional X-ray sources present in the uncalibrated 
field centered 3 arcmin N of the nucleus. 
Column 1 gives our identification number, followed by RA and DEC (J2000)
as listed by Kraft et al. (2001).
Columns 4 and 5 give the 
X-ray luminosities measured by Kraft et al. (2001).  

While there is significant overlap with
the central field, Table 2 lists the additional sources, for which no 
photometry can be obtained. Nine of these bright X-ray sources show clear
optical counterparts,
and from the extended morphology we have classified potential globular
clusters. 
Thus, the selection of "GC?" and "FC" sources for the sources in Table 2
is only based on the PSF sizes, and
these candidates have to be confirmed with accurate colors. 

Comparing with the previous papers, the present work represents 
more than a factor of 4 improvement in matching the optical counterparts.
While some 10\% of the sources are identified by Kraft et al. (2001) in the
large field they studied, we identify 60\% of the CHANDRA sources in the
inner regions of NGC5128. 

\section{Discrimination of Other Sources}

Discrete X-ray sources in distant galaxies can be X-ray binaries or supernova
remnants (Fabbiano 1989).
In order to discriminate globular clusters from other sources
like background emission line galaxies, or NGC5128 SN remnants and WR stars
(or young WR clusters),
we use a continuum-subtracted H$\alpha+[N~{\scriptsize II}]$ emission image.
As Fabbiano \& Trincheri (1987) have pointed out, there 
is a close relationship between the X-ray emission and the blue stellar 
population. The X-ray surface brightness generally follows the optical 
light distribution. It is also well known that H$\alpha$ emission is a 
good tracer of current star formation. Therefore in the case of NGC 5128, 
we can discard those X-ray sources associated also to H$\alpha$ emission 
because they most probably correspond to young clusters or SN remnants.

The narrow-band imaging centered on NGC5128 was performed on May 26, 2001 at
the 0.9m telescope at Cerro Tololo Inter-American Observatory (CTIO).
The 2048$\times$2048 Tek \#3 CCD 
yielded a field of view of 13$\times$13
arcmin$^2$ with an image scale of 0.8 arcsec pix$^{-1}$ after a 2$\times$2
on-line pixel binning. 
We obtained 3$\times$600 second narrow-band images with the 
interference filter ($\lambda_c = 6598$ \AA ; $\Delta\lambda_{FWHM} = 69$ \AA)
isolating the spectral region characterized by the redshifted H$\alpha$ and
$[N~{\scriptsize II}]$ (6548, 6583 \AA) emission lines. To subtract the
stellar continuum we use 3$\times$60 second $R$-band images 
taken with the broad-band filter Rtek2 ($\lambda_c=6425$ \AA ; 
$\Delta\lambda_{FWHM} = 1500$ \AA). Different dome and sky flat-field 
exposures were taken for each of the two filters.
Data reduction was carried out using IRAF. 
The final continuum-free image of NGC5128 was obtained by subtracting the
combined $R$-band image, suitably scaled, from the combined narrow-band image.
The mean scale factor for the continuum image was estimated by
comparing the intensity of the field stars in the two bandpasses.

The final H$_{\alpha}$ image shows striking similarity to the map of all
point sources shown in Figure 2. The gas emission is located following
the disk blue sources in the region.
A dozen sources show H$_{\alpha}$ emission counterparts in the inner dust disk.
These are identified in the last column of Table 1. None of the
sources listed in Table 2 appear to have H$_{\alpha}$ emission.

Regarding the background sources,
the surface density of X-ray background sources has been estimated by
Giacconi et al. (2001). In the $6\farcm8 \times 6\farcm8$
field of the VLT there should be about 12 background X-ray sources.
A total of 47 of the sources listed in Table 1 do not have optical counterparts
down to $B=25$.  Some of these sources without optical counterparts
can well be background sources such as distant AGN and QSO. 
Four of them (sources 6, 21, 32, and 54) have a matching faint H$\alpha$ source, in 
spite of not showing any counterparts in the broad-band images.

We have also inspected the X-ray sources with optical counterparts individually,
and none of them looks like an obvious galaxy. Judging by their sizes and 
ellipticities, most resolved sources are globular clusters. 

Objects 19, 39, and 42 have broad-band colors consistent with 
globular clusters, but they show H$_{\alpha}$ emission, and are
thus discarded from the globular cluster candidate list.
However, note that some globular clusters may exhibit emission lines
(e.g. Minniti \& Rejkuba 2002).

\section{Foreground Milky Way Sources}

One caveat worth mentioning is that NGC5128 lies at an intermediate Galactic
latitude, and a number of foreground Galactic sources are expected.
X-ray emitting stars can be chromospherically active stars such as
RS~CVn variables, active late-type stars of M-Me types,
cataclysmic variables, interacting binaries, etc. (e.g.
Pellegrini \& Fabbiano 1994).

We identify  a few of these bright foreground stars, as they are not
resolved like most bright NGC5128 globular clusters. One source 
is in the bled column of a saturated star, another one of these
stars is saturated
and two other 
(sources 85 and 92)
have colors and magnitudes consistent
with M-type dwarfs.  Sources number 85 and 92 are the two reddest objects 
listed in Table 1 and plotted in Figure 4. 
Their colors are
consistent with either being M-type stars or very reddened globular clusters.
We exclude them from our final sample of globular clusters
listed in Table 3.

The distinction between Milky Way stars and distant globular clusters, 
however, is more difficult for fainter magnitudes. For magnitudes 
$V<22$ we rely solely on the colors of the sources to distinguish 
stars from globular clusters.

\section{ X-Ray Sources in Globular Clusters: NGC5128 $vs$ the Milky Way}

The deep high resolution VLT images allow the identification of new globular clusters
in the inner regions of NGC5128, a task otherwise difficult due to reddening
and high surface brightness of the underlying galaxy.

Table 3 lists all the CHANDRA point sources that can be associated with 
NGC5128 globular clusters.
There are in total 33 sources, 29 of which are located in our field
(see Table 1), plus 4 that are outside our field: 2 from Harris
 et al. (1992) that Kraft et al. (2001) associated with X-ray sources, and 
2 from Rejkuba (2001) that we find that are matched in positions with the
CHANDRA X-ray point sources. Harris et al. (1992) give radial velocities,
magnitudes and colors for G176 and G284, confirming that they are
true NGC5128 globular clusters. Rejkuba (2001) gives sizes, colors and
magnitudes for clusters clusters f1.16 and f2.81, also confirming 
them as globular clusters of this galaxy. 

In Table 3, column 1 gives our identification number, followed by RA and DEC 
(J2000) as listed by Kraft et al. (2001).  Columns 4 and 5 give the 
X-ray luminosities measured by Kraft et al. (2001). For 
consistency they have not been changed, but we note that
these X-ray luminosisites should be increased by
24\% to bring them to our preferred distance of 
$D=3.9$ Mpc instead of their adopted distance of $D=3.5$ Mpc 
(Rejkuba 2003).  Columns 6 to 8 list the
$BVI$ magnitudes from the VLT images.  
Column 9 tells
whether the globular cluster has been confirmed spectroscopically.
Finally, column 10 gives the sources of other globular cluster identifications.

About 10\% of the Milky Way globular clusters harbor bright
X-ray sources, with $L_X>10^{36}$ erg sec$^{-1}$. The globular cluster
system of NGC5128 contains about 1700 clusters, and we expect 150-200
of them to harbor bright X-ray sources. 
Deeper CHANDRA observations of NGC5128 would
be useful to identify fainter globular cluster sources, and to monitor their
variability.

There are relatively more globular cluster sources in NGC5128 that
have luminosities above $L_X>10^{37}$ erg sec$^{-1}$ (Kraft et al. 2001). 
About 2/3 of NGC5128 globular clusters are that bright, compared with 1/3 in
M31 and 1/10 in the Milky Way.

Only 1 out of 150 of the Milky Way globular clusters has an X-ray source 
as bright as $L_X>10^{37}$ erg sec$^{-1}$ like the sources detected in NGC5128, 
although we note that the luminosities quoted by Verbunt
et al. (1995) are for a softer energy band, and they miss sources like NGC6440, 
plus the fact that there are a few variable sources in the Milky Way that have 
shown X-ray luminosities in excess of $10^{37}$ erg/sec.
Taking the value from Verbunt et al. (1995),
we should then expect 10-12 sources in NGC5128 with $L_x>10^{37}$ erg sec$^{-1}$,
but there are 24 in total, in the central region studied
here alone (when multiple epoch observations are taken, we
use the mean for this comparison). This implies a frequency of X-ray bright
globular clusters that is at least two times higher
than that of our Galaxy. This result agrees with previous findings
for M31 and other galaxies (Trinchieri \& Fabbiano 1991, Primini et al. 1993,
Di Stefano et al. 2002, Sarazin et al. 2001,
Angelini et al. 2001, Finoguenov \& Jones 2002,).
Note that the contrast would be increased using a distance of $D=3.9$ Mpc
(Rejkuba 2003), because the luminosities should be increased
by 24\% with respect to the values given by Kraft et al. (2001).

This lower limit to the frequency of X-ray bright globular clusters is well
defined, but the upper limit is not.
A crude scaling of the total number of sources examined here comparing with 
the total globular cluster population allows us to predict that there
should be $>100$ globular clusters in NGC5128 hosting bright X-ray sources,
a frequency up to $10\times$ higher than that of the Milky Way.

\section{ X-Ray Sources in Globular Clusters: NGC5128 $vs$ M31}

Aside from our Galaxy, the best studied population of
globular cluster X-ray sources is that of M31 (see Trinchieri \& Fabbiano
1991, Primini et al. 1993, Di Stefano et al. 2002).
In comparison, there are 10\% more X-ray binaries with luminosities
$L_X>10^{37}$ erg sec$^{-1}$ in NGC5128 per unit optical luminosity 
than in the bulge of M31, and $25\%$ more than
in the whole M31 (Kraft et al. 2001). This proportion is also larger if we
count only the globular cluster X-ray sources.

Di Stefano et al. (2002) found that, first, the peak X-ray luminosity of 
M31 globular clusters is higher by $10\times$ than that of the Milky Way, and
second, that a larger fraction of globular cluster X-ray sources have 
luminosities with $L_X>10^{37}$ erg sec$^{-1}$. In NGC5128 we find similar
results.

Di Stefano et al. (2002) discuss the brightest globular cluster X-ray source
in M31 with $L_X=2-6\times 10^{38}$ erg sec$^{-1}$, 
which is located in the cluster Bo 375. Among the sources in the
fields studied here, there is one source as bright in X-rays as Bo 375 in M31:
source 10 in Table 1.  This bright NGC5128 source that reaches
$L_X=2.4\times 10^{39}$ erg sec$^{-1}$ in one X-ray observation drops to
$L_X<2\times 10^{36}$ erg sec$^{-1}$ in the other X-ray observation of
Kraft et al. (2001) -adopting $D=3.9$ Mpc. 
This very bright and variable source has no optical
counterpart in the VLT images and has no H$\alpha$ emission, and it is
deffinitely not associated with a globular cluster like Bo 375 in M31. 

So far Bo 375 continues
to hold the record as the brightest X-ray source in a spectroscopically
confirmed globular cluster, although
NGC 1399 and NGC 4472 appear to have brighter Xray sources with
optical counterparts (Angelini et al. 2001, Irwin et al. 2002).
This means that the peak X-ray luminosity function of globular
clusters in M31 is still brighter than that of NGC5128, despite the larger
globular cluster system of this gE galaxy.

As discussed by Di Stefano et al. (2002) in the case of M31, the most luminous
X-ray sources in globular clusters can be multiple sources instead of
black hole accretors. Assuming Poisson statistics,
the probability of observing a multiple source increases
with the number N of sampled objects.
In our case, taking the probability of finding an X-ray source in a
globular cluster as $p=0.1$ (from the Milky Way), leads to the prediction that
$N=3$ globular cluster X-ray sources in NGC5128 would be composed of two 
sources, while the number of triple sources is negligible. 
Thus, a few of the bright sources listed in Table 1
could be the superposition of two LMXBs
inhabiting a globular cluster.
The dynamical implications of these are interesting and would not be discussed
here. However, we note that Miller \& Hamilton (2002) have recently suggested
that four-body interactions can lead to black hole formation in globular clusters.

Most clusters in the inner regions of NGC5128 are distributed within
the dust lane, and are relatively young objects. In contrast,
the majority of the X-ray sources with optical counterparts
are associated with old globular clusters. Only 2-3 of them can be
associated with younger, bluer clusters according to the color-magnitude
diagrams (Figure 3).
The differences between these populations is
also seen in Figure 2, where the distribution of X-ray sources does
not follow the inner dust lane. While centrally concentrated, these
X-ray point sources are uniformly distributed about the center. This
is expected if they are preferentially associated with old stellar
populations such as globular clusters.

\section{ X-Ray Sources in Globular Clusters: NGC5128 $vs$ Distant Elliptical Galaxies}

What is the frequency of bright X-ray sources in globular clusters in galaxies?
There are in the literature different values for this frequency, partly
due to the different X-ray flux limits.
On the high end, Angelini et al. (2001) measure a frequency of 
70\%  for the giant elliptical galaxy NGC1399, and Kundu et al. (2002)
measure a frequency of 40\% for NGC4472.  Intermediate values are those of
NGC1553 with $\sim$20\% (Blanton et al.\ 2001),
NGC4697 with a lower limit of 25\% (Sarazin et al.\ 2000, 2001), and
M31 with  20\% (Di Stefano et al.\ 2002).
On the very low end there were NGC5128, where the 9 clusters identified by
Kraft et al. (2001) out of 246 sources implied a frequency of
$3-4\%$,  and off course the Milky Way, with a frequency of about $10\%$ of 
$L_X>10^{36}$ erg sec$^{-1}$ sources (Verbunt et al. 1995).
Based on all these data,
it is not obvious to decide how this frequency changes with galaxy type. 
We stress the importance of systematically identifying optical counterparts
of CHANDRA X-ray point sources in distant galaxies using high quality data 
such as provided by the VLT or HST.
The present work increases $10$ times the frequency for 
NGC5128 X-ray sources in globular clusters to 30\%, allowing a better
agreement among these galaxies. 

As noted by Kundu et al. (2002), the fraction of globular cluster hosting
bright LMXBs is similar for different galaxies, of the order of 1\%$-$4\%.
In this respect, we find that NGC5128 appears to be no different than
other galaxies.

The question of the frequency of X-ray globular clusters is relevant for
LMXB formation models.  White at al. (2002) suggest that all LMXBs may 
form in globular clusters, some of which are later ejected from the
globular clusters by supernovae kick velocities, by stellar interactions,
or by the evaporation of the clusters due to tidal effects (Sarazin et al. 
2001).  In this scenario, the total LMXB population must correlate with the
globular cluster systems of galaxies. 
It should be noted though that LMXB populations need to be determined
in a larger number of galaxies in order to check this picture
(Sarazin et al. 2001).
This viewpoint has problems according to Finoguenov \& Jones (2002)
on the basis of the low frequency of X-ray globular clusters measured in
galaxies such as NGC5128. 
Our new frequency of $30\%$ for this galaxy lends support to
the idea that all LMXBs form in globulars.
Interestingly, the LMXB
population could give us a hint about the original population of globular
clusters in galaxies, of which we only see the survivors. This, however, 
depends on the mean LMXB lifetimes.

Figure 5 shows the X-ray luminosity distribution of all sources compared
with those located in NGC5128 globular clusters.  This figure plots each object
twice, corresponding to both epochs of observations of Kraft. et al. (2001).
We note that plotting the averages does not change the graph significantly.
When comparing this luminosity distribution
to more distant galaxies, we will restrict the sample to
$L_x>10^{37}$ erg sec$^{-1}$, keeping in mind that this limit moves
depending on the distance adopted for NGC5128. 
~From Table 1 and Figure 5, it can be seen that 50\% of all sources with 
$L_x>10^{37}$ erg sec$^{-1}$
are globular clusters. 
We note that the association of
X-ray point sources to globular clusters in more distant galaxies is tricky.
On one hand, more optical point sources are seen within the matching
radius, leading to false identifications. On the other hand, only the
brighter X-ray sources and the brighter optical sources
will be detected, leading to fewer identifications.
It is difficult to model these competing effects for galaxies with different
structures and globular cluster systems, in the presence of varying 
backgrounds and dust, in order to correctly
interpret the optical identifications of X-ray sources in globular
clusters of very distant galaxies. That is also why a census
of globular cluster X-ray sources in NGC5128 is important.

\section{Metallicities and Luminosities of X-ray Globular Clusters}

Do bright X-ray point sources reside preferentially in metal-rich globular 
clusters in NGC5128?  The globular clusters with X-ray sources 
in NGC5128 have preferentially red colors,
as seen in Figure 3. For the globular cluster X-ray sources in NGC1399
and NGC4472 there is a tendency to be redder than GCs in general
(Angellini et al. 2001, and Kundu et al. 2002).
This tendency is confirmed in NGC5128 here.  
 
Figure 6 shows the color distribution of sources with $I<22$,
which are mostly globular clusters,
present in the VLT images outside the dust lane region, compared
with the X-ray globular clusters.  We restrict the
comparison to the brighter magnitudes to avoid contamination
of fainter sources. 
 This color distribution is bimodal as have been found
in the outer NGC5128 regions (Zepf \& Ashman 1993, Rejkuba 2001), 
as well as in the inner regions of this galaxy (Minniti et al. 1996, 
Held 2002). The bimodal color distribution of globular clusters
in galaxies is interpreted as a bimodal metallicity distribution. In NGC5128,
the most metal-rich clusters appear to be more concentrated towards the inner
regions than the more metal-poor clusters.  The color distribution
of all globular clusters in the VLT field is different than
the color distribution of globular clusters containing
X-ray sources listed in Table 3 (this difference is significant 
at the 95\% confidence according to a Kolmovgorov-Smirnoff test).
The X-ray globular clusters are preferentially red, very few blue clusters
(with $V-I<1.1$) harbor X-ray sources.  Adopting the traditional
view that clusters with redder colors are more metal-rich,
we conclude that globular cluster X-ray sources tend to be preferentially 
located in metal-rich clusters in NGC5128.

Do X-ray sources reside preferentially in bright globular clusters? 
For NGC1399, Angelini et al. (2001) found that the clusters with X-ray sources
are brighter in general than the rest of the globular cluster population.  
A similar trend was found by Kundu et al. (2002) for NGC4472.

Figure 7 shows the magnitude distribution of globular cluster X-ray sources in 
NGC5128 from Table 3 compared with the NGC5128 globular cluster
luminosity function of Rejkuba (2001).
The peak of the magnitude distribution of X-ray globular clusters
is located at $V=19$ or $M_V=-9.5$.
It is clear that, although a wide magnitude distribution is present, the
globular clusters with X-ray sources are about 2 magnitudes brighter in the
mean than the rest of the population.
However, the dependence showed here  may reflect the fact that the Rejkuba
data are be deeper.

This confirms and strengthens the results of Angelini et al. (2001) and
Kundu et al. (2002), in that
bright X-ray sources preferentially belong in bright globular clusters.
Our faintest source, with 
$L_x=1.5\times 10^{36}$ erg sec$^{-1}$, is about 10 times fainter than the 
sources studied by Angelini et al. (2001) and Kundu et al. (2002), so these
results hold even for fainter X-ray fluxes.

The globular cluster X-ray sources in our galaxy tend to be located in 
brighter clusters. There are only 3 Milky Way globular clusters
with X-ray sources that are fainter than the peak of the globular
cluster luminosity function, while 9 of them are brighter (Verbunt et al. 1995).
In NGC5128, as in the Milky Way, it appears that the
X-ray sources preferentially reside in bright clusters. 
Di Stefano et al. (2002) also found that M31 X-ray globular clusters are
brighter in the $V$-band (and possibly more massive) 
than the Milky Way X-ray globulars. 

Do the brightest LMXBs reside in the most massive clusters?
There is no trend for brighter globular clusters to have brighter X-ray
sources in NGC5128 (see Figure 8). 
The luminosities of the sources within and outside
globular clusters is not significantly different, although there seem to
be more naked sources that are brighter than $L_x=10^{38}$ erg sec$^{-1}$
(as seen in Figure 5). However, the number of sources is small to draw
any conclusion.

\section{High Mass Accretors and Black Holes}

The inner dust lane of NGC5128 contains a large number of blue objects,
which are bright stars and clusters.  We can also try to identify
potential optical counterparts of high mass X-ray binaries (HMXBs) and
high mass-rate accretors, including black holes. 
HMXBs are accreting binaries where 
the donor is an O or B-type star, and are in general less numerous
than LMXBs. 

Figure 8 shows the X-ray luminosities of all matched sources $vs$
$B$-band magnitude and $B-V$ color, respectively. For each source, two
luminosities are plotted corresponding to the two different epochs of
CHANDRA observations.
The different numbers of sources in these are due to the 
different completeness/saturation limits in the $B$ and $V$-bands.
No clear trends are
seen either with colors or with magnitudes in these figures. The only source
that stands out is the very blue source (\# 15) in Figure 8, but we note that
many other faint objects are detected in the $B$-band and not  in
$V$ or $I$, being in consequence bluer than this source.
The blue sources are mostly located within the dust lane, where recently
formed stars and clusters are seen. 

The brightest X-ray sources with luminosities exceeding $10^{37}$ erg sec$^{-1}$ are
LMXBs containing a donor with $M<1M_{\odot}$ and an accretor
neutron star. These LMXBs have high accretion rates, as large as 
$10^{-8}$ M$_{\odot}$ yr$^{-1}$ (Piro \& Bildsten 2002).
They are short lived compared to the ages of globular clusters and
Piro \& Bildsten (2002) suggested that they are long-term transients.
They argue that, in isolation, the optical counterpart of 
a typical source like Cygnus X-2 would have 
$M_V=-2.0$, and would be very blue, with
$B-V\approx 0$, $U-B\approx -1$. For NGC5128, this corresponds
to blue sources with 
$V\approx 26$, beyond the magnitude limit of
our dataset.

There are 11 sources with clear optical counterparts in the $B$-band, but
without colors. Five of them are associated with globular clusters that
have saturated $V$ and $I$ magnitudes. The other 6
are not associated with globular clusters, being too faint and too blue 
($B-V<0$) to be globular clusters. Three of these isolated LMXBs are very variable
(a factor of $2\times$ or more)
between the two CHANDRA observations. 

We now concentrate on the brightest X-ray sources in our field.
There are 9 bright sources with $L_X>10^{38}$ erg sec$^{-1}$ at least in one
of the epochs observed by Kraft et al. (2001): four of them have no optical
counterparts in the VLT images, four of them are only seen in the B images,
implying very blue colors ($B-V<0$), and only one of them appears to be
a globular cluster source, although it shows faint H$\alpha$ emission. 
Of these 9 sources, 6 of them have varied 
by $2\times$ or more between the two CHANDRA observations. 

Source 16 of Table 1 is CXOU J132526.4-430054 of Kraft et al.  (2001). 
They found that this very bright X-ray source located in the dust lane
20" NW of the nucleus has a hard
spectrum, atypical of an AGN or a LMXB. They raise the possibility of a
HMXB that is heavily obscured. We find no $BVIH{\alpha}$ counterpart
for this source down to $B=25$, consistent with the obscured HMXB hypothesis.

Source 72 of Table 1 is CXOU J132519.9-430317 of Kraft et al. (2001).
This source at 2.5 arcmin SW of the nucleus 
is a bright recurrent transient that is highly variable.
Its peak luminosity exceeds the Eddington limit of X-ray binaries.
No $BVIH{\alpha}$ counterpart is found down to $B=25$. 
Kraft et al. (2001) argue that the decrease in luminosity by $500\times$
cannot be due to a background AGN, and the lack of an optical counterpart
reinforces this hypothesis.

There is one blue source (number 15 in Table 1)
with $V=23.1$ and $B-V=-0.4$ that also has H$\alpha$ emission, which would be
interesting to follow up searching for variability. It can be seen as
the bluest X-ray source in the left panel of Figure 3, and of
Figure 8.  This source is located close to the nucleus of the galaxy, 
in a very dusty region.  It is probably too faint
to be the nucleus of the galaxy that NGC5128 accreted in the past.

Finally, 
of the bright sources associated with globular clusters, it is interesting
to discuss the possible presence of black hole accretors. These should be
very energetic, largely variable, and short lived. None of the X-ray sources 
in the Milky Way globular clusters appears to be a black hole accretor. However,
only one Galactic cluster has $L_x>10^{37}$ erg sec$^{-1}$. NGC5128 has a larger
globular cluster system where these rare stages of stellar evolution can
be probed. In this galaxy we find 24 globular clusters with such high 
luminosities, as can be seen in Table 3. 

Variable sources are also common among the X-ray globular clusters.
Table 3 shows that there are 14 globular cluster X-ray sources that varied by a 
factor of 2 or more between the two CHANDRA observations of Kraft et al. (2001).
This includes source 17, which is
the brightest globular cluster source in NGC5128 that we identify,
reaching a luminosity of 
$L_x=1.66\times 10^{38}$ erg sec$^{-1}$.
This is just short of the Eddington luminosity of a 1.4 $M_{\odot}$ neutron
star, $L_{x Edd}=2\times 10^{38}$ erg sec$^{-1}$ (White et al. 2002).
Thus, none of our globular cluster X-ray sources in NGC5128 appears to be 
an accreting black hole.

\section{Conclusions}

NGC5128 is a very special galaxy for the study of globular cluster
LMXBs.
Its proximity allows to detect X-ray sources as faint as  $L_X
\approx 10^{36}$ ergs s$^{-1}$ (Kraft et al. 2001), 
and at the same time it allows to find
its faintest globular clusters (Rejkuba 2001). Since its globular cluster
system is so large ($>10\times$ that of the Milky Way), we can find 
objects that trace rare stages of stellar evolution like LMXBs. 
This galaxy can be used as a tool to link our knowledge of local
globular cluster X-ray sources with those of more distant giant galaxies.
The task is not without difficulties, as NGC5128 has a thick dust lane, and
very many recently formed clusters in the inner regions.

We identify 33 globular cluster X-ray sources in the inner regions of NGC5128,
including 23 that are newly discovered. Of these, 20 have been classified as
globular clusters on the basis
of their sizes, magnitudes and colors, and 3 only on the basis of their sizes.
The reliability of the
identification has been confirmed spectroscopically for 11 of them.

Thus, this work 
confirms that a large fraction of the bright X-ray point sources  found in
elliptical galaxies are associated with globular clusters, 
likely being X-ray binaries. 
For the inner regions of
NGC5128, 30\% of all X-ray sources with $L_X>2\times 10^{36}$ ergs s$^{-1}$
are located within globular clusters.

Even though the color distribution of globular clusters
in the inner regions is bimodal with a wide color range, most of the globular 
clusters with X-ray sources have red colors, with $1.0<V-I<1.5$.
Assuming that clusters with redder integrated colors are more metal-rich,
we conclude that globular cluster X-ray sources tend to be preferentially 
located in metal-rich clusters.  

These clusters also tend to be brighter in the
mean than the rest of the population, 
although a wide range in magnitudes is present ($-12<M_V<-6$). 
Assuming $M/L=3$, typical for Milky Way globulars, this means that
LMXBs are preferentially located in more massive clusters.

In the Milky Way, we are dealing with small number statistics of
LMXBs in globular clusters. However, it appears that
the population of NGC5128 globular cluster
X-ray binaries is brighter than the X-ray binaries
found in globular clusters of the Milky Way. In our galaxy
there is only one globular cluster with  $L_X> 10^{37}$ ergs s$^{-1}$,
while we found 24 such sources in NGC5128. Only a dozen where expected
in the whole of NGC5128 by just
scaling the sizes of the globular cluster systems of these galaxies.
However, none of the X-ray globular clusters identified here
appears to harbor an accreting black hole. 

We also identify the optical counterparts of 
29 X-ray point sources that are not NGC5128 globular clusters.
Finally, 53 X-ray sources (48\% of the population)
do not have any optical counterparts
down to the faintest magnitude limits $B=25$ (a dozen background sources
are expected in the field).

The challenge now would be to combine the deepest possible CHANDRA
images with the highest quality optical images to do a complete systematic
survey of LMXBs in a wide variety of
elliptical galaxies, in order to test the hypothesis that all LMXBs form in
globular clusters, recently proposed by White et al. (2002).

\acknowledgments{
We are grateful with the IATE Group at C\'ordoba Observatory in Argentina
for their hospitality while this paper was being written.
DM is supported by Fondap Center for Astrophysics 1500003.
 }

\clearpage
\begin{figure}
\plotone{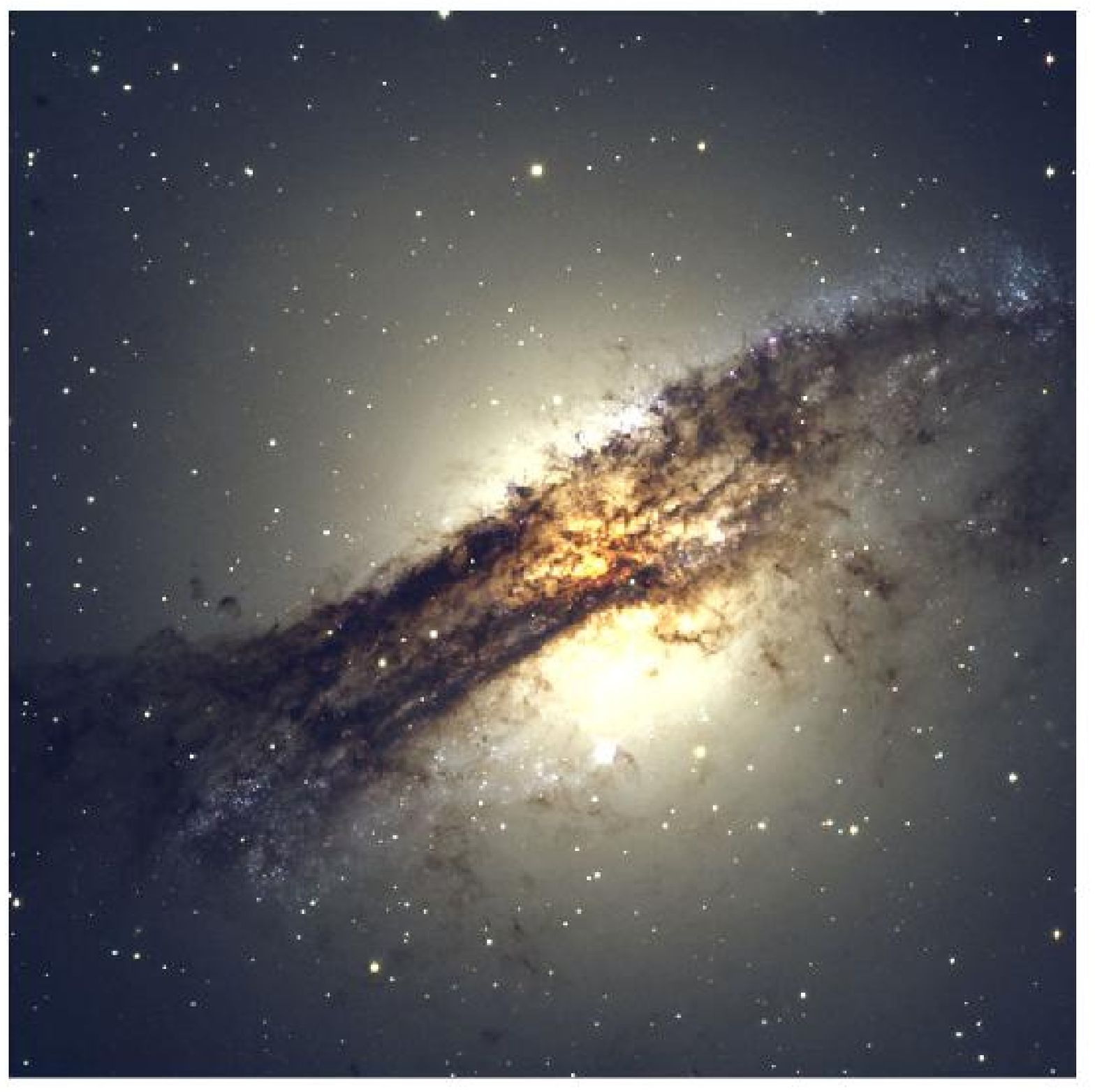}
\caption{
Color VLT image of NGC5128
from ESO Press Release Photo 05b-00. 
The field of view is 
$6\farcm8 \times 6\farcm8$;
North is up, and East to the left.
}
\end{figure}

\clearpage
\begin{figure}
\plotone{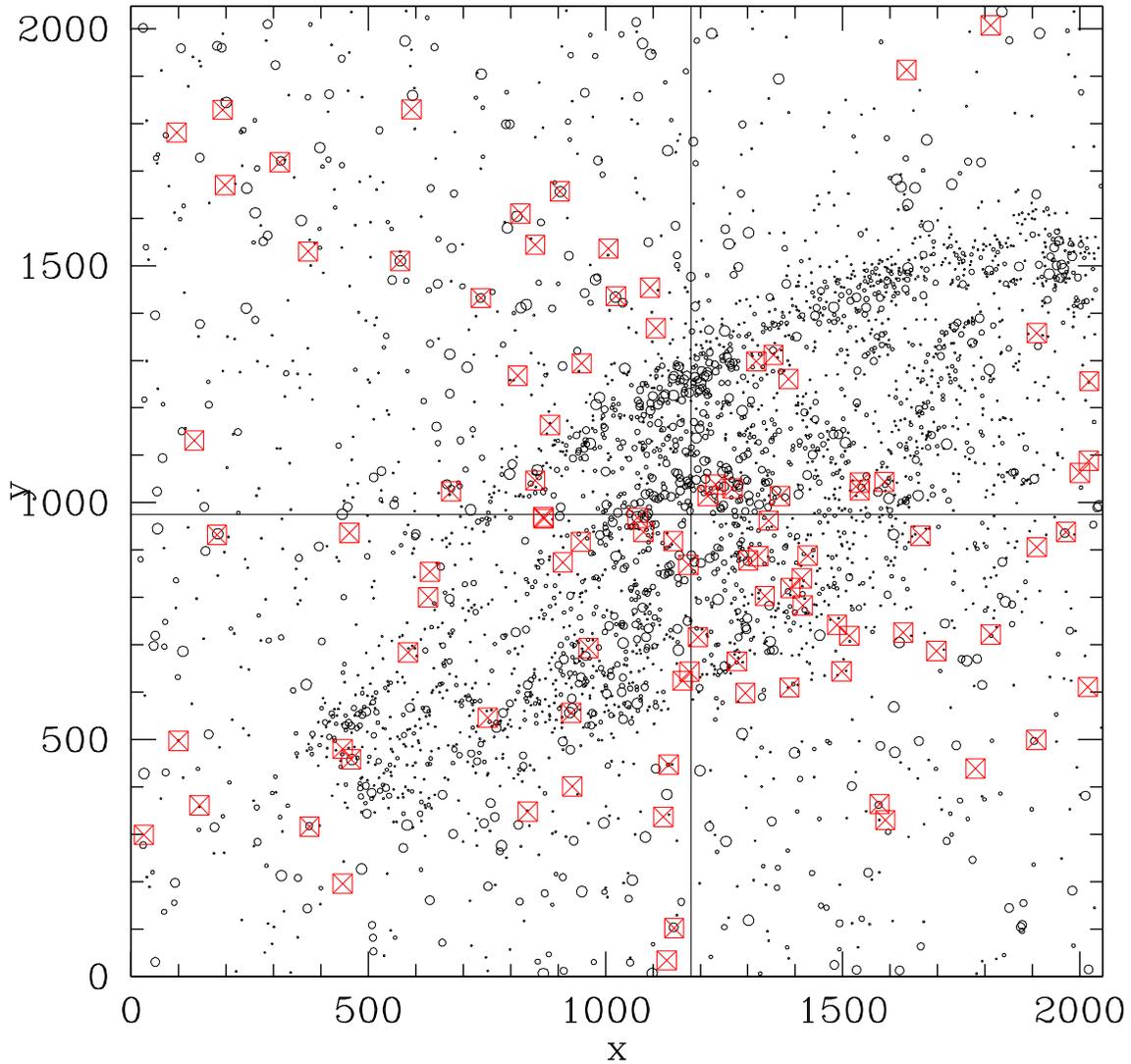}
\caption{
Location of the of the point sources in the VLT images of the
inner 
$6\farcm8 \times 6\farcm8$
of NGC5128. The X-Y scales are in pixels;
North is up, and East to the left; the center of the galaxy is indicated.
The sizes of the circles are proportional to the $B$-band magnitudes.
The positions of the CHANDRA point sources from Kraft et al. (2001) 
are indicated with crossed squares. Beware that saturated objects ($B<19$)
are not plotted.
}
\end{figure}

\begin{figure}
\plotone{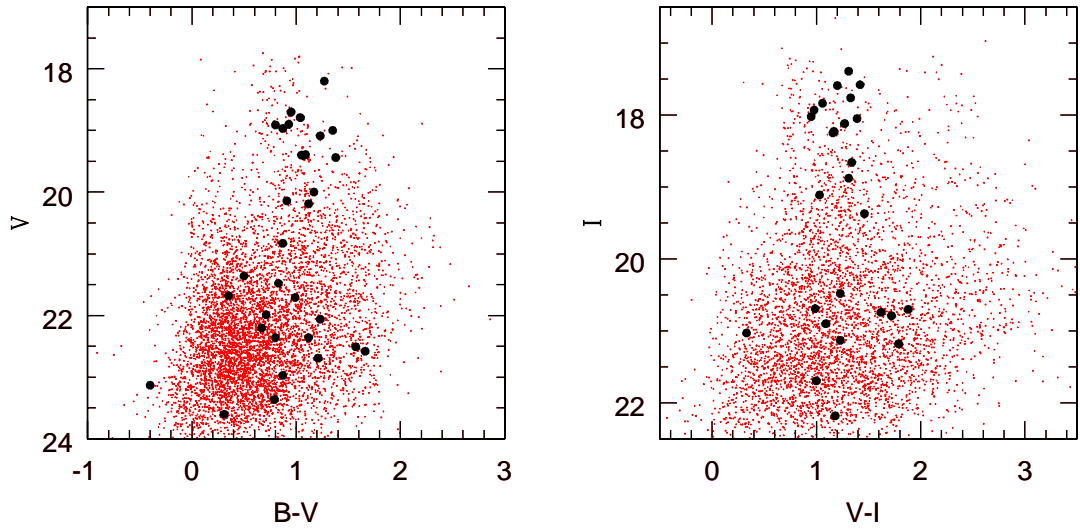}
\caption{
Color-magnitude diagrams of point sources present in
the VLT images with best photometric data ($\sigma_V < 0.1$).
These include all sources, even the ones present in the inner dust lane,
so strong differential reddening is expected.
The larger black circles indicate the X-ray sources that
have been matched to optical counterparts.
}
\end{figure}


\begin{figure}
\plotone{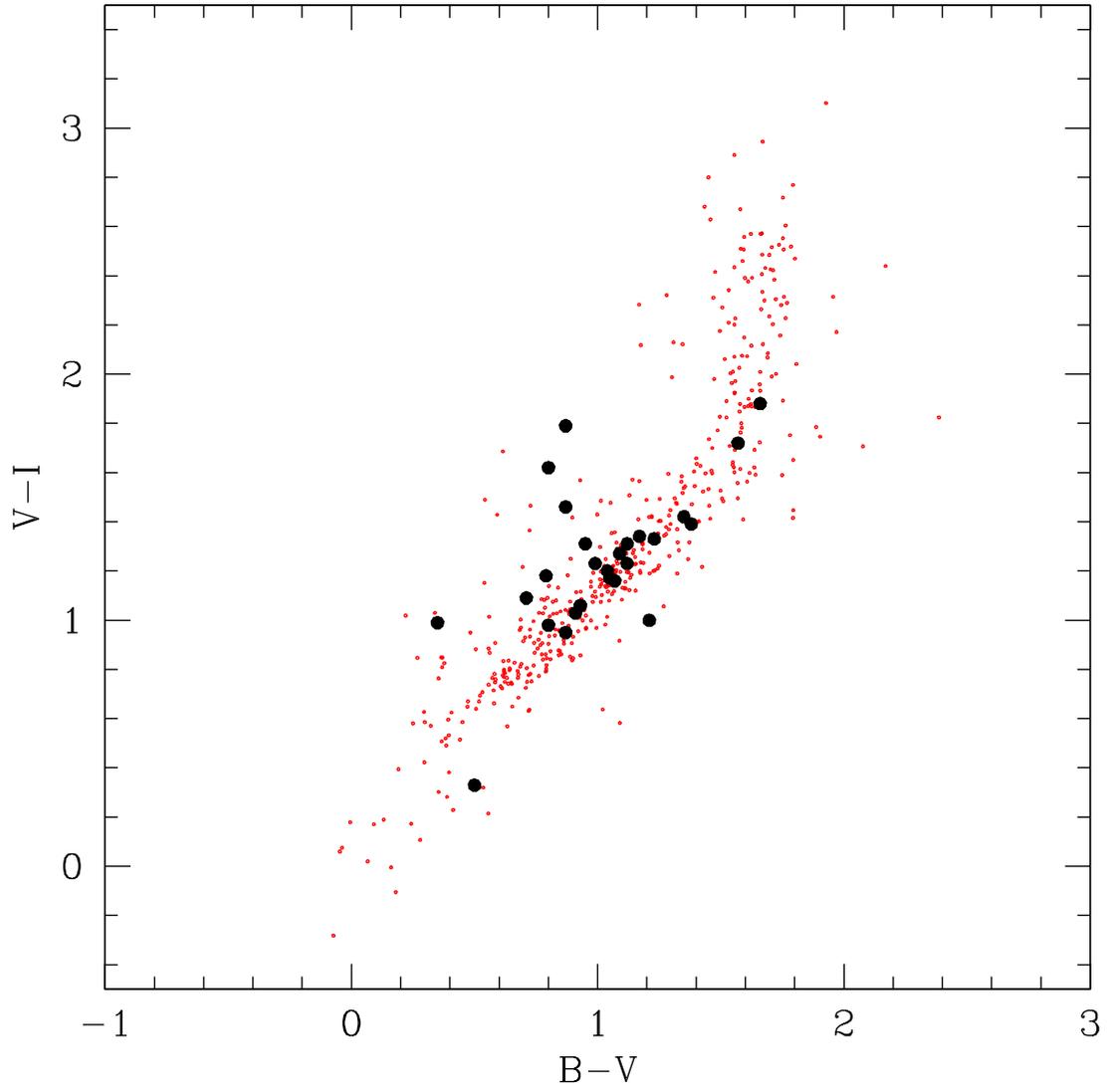}
\caption{
Color-color diagram of point sources present in
the VLT images outside the dust lane region where most younger sources
are concentrated.  The larger black circles indicate the X-ray sources that
have been matched to optical counterparts.
}
\end{figure}

\begin{figure}
\plotone{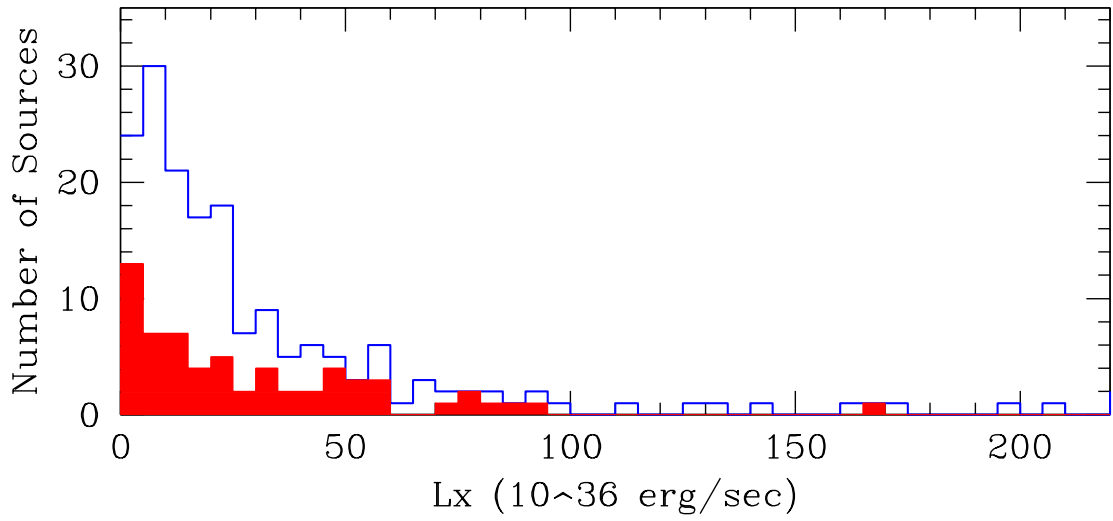}
\caption{
X-ray luminosity distribution for all point sources in the field (open 
histogram) compared with those sources in globular clusters (solid histogram).
Two counts are plotted for each source 
when two fluxes are given by Kraft et al. (2001). 
}
\end{figure}

\begin{figure}
\plotone{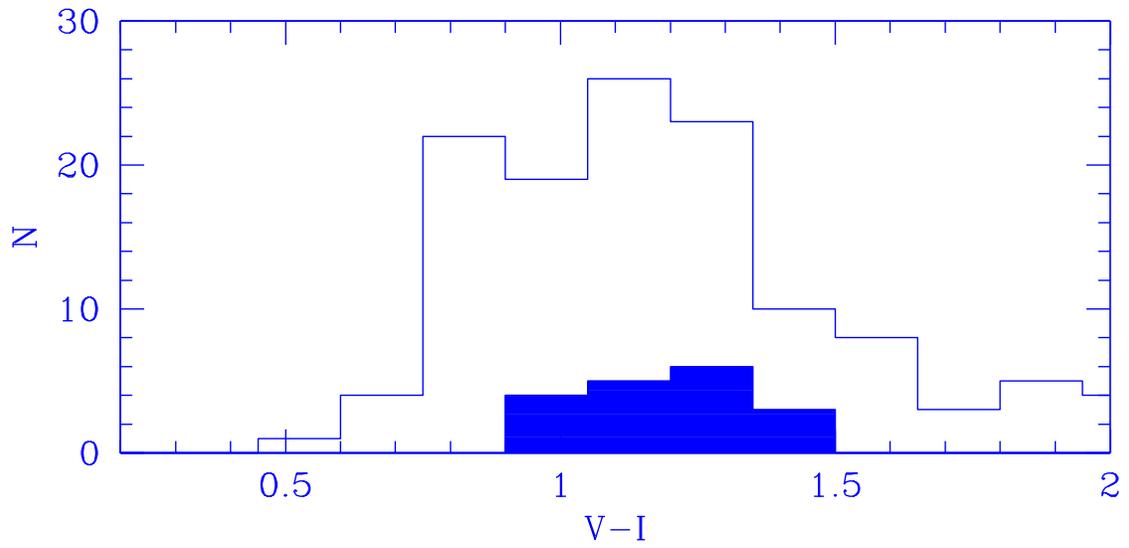}
\caption{
Color distribution of globular cluster X-ray sources in 
NGC5128 from Table 3 (solid histogram), compared with the
color distribution of sources with $I<22$
(empty histogram)
present in the VLT images outside the dust lane region
 (mostly globular clusters, but some may
be foreground stars or background galaxies).
}
\end{figure}

\begin{figure}
\plotone{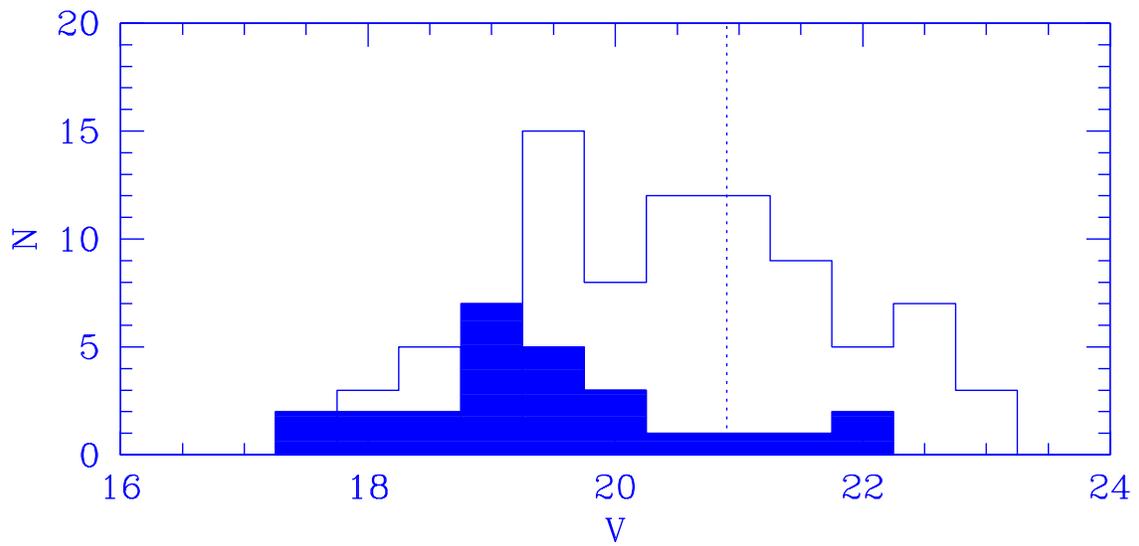}
\caption{
Magnitude distribution of globular cluster X-ray sources in 
NGC5128 from Table 3 (solid histogram). The peak of this magnitude distribution 
is 2 magnitudes brighter than the
expected peak of the NGC5128 globular cluster luminosity function, located at
$M_V=-7.4$, or $V=20.9$
(dotted line). The globular cluster luminosity function of Rejkuba (2001)
is plotted for comparison (empty histogram).
Note that 3 saturated globular clusters (with $V<18$) are missing. 
}
\end{figure}

\begin{figure}
\plotone{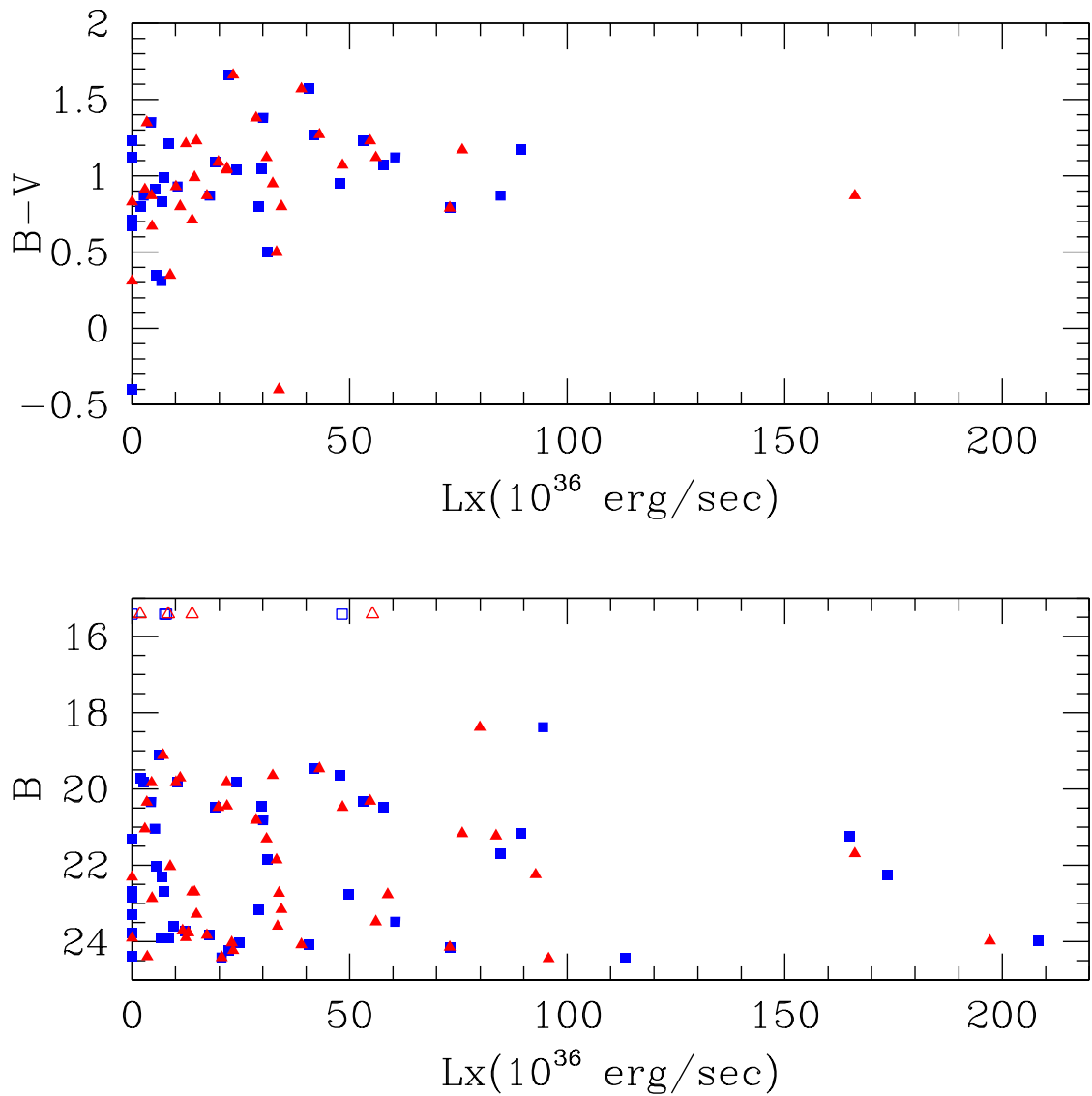}
\caption{
Top: X-ray flux $vs$ B-V color for all point sources matched in the field.
Two points are plotted for each source (a square and a triangle)
when two fluxes are given by Kraft et al.  (2001). 
No trend of X-ray luminosity with optical color is seen. 
Bottom: X-ray flux $vs$ B-band magnitude for all point sources matched 
in the field.  Open symbols at the top of the plot
depict B-magnitude upper limits, corresponding to bright optical counterparts
that are saturated objects in our VLT images. No trend of X-ray luminosity 
with $B$-band magnitude is seen.
}
\end{figure}

\begin{planotable}{llllllllllllll}
\small
\footnotesize
\tablewidth{0pt}
\scriptsize
\tablecaption{Optical ID of NGC5128 X-Ray Point Sources}
\tablehead{
\multicolumn{1}{c}{MRF}&
\multicolumn{1}{c}{RA}&
\multicolumn{1}{c}{DEC}&
\multicolumn{1}{c}{$L_X$316}&
\multicolumn{1}{c}{$L_X$962}&
\multicolumn{1}{c}{X}&
\multicolumn{1}{c}{Y}&
\multicolumn{1}{c}{B}&
\multicolumn{1}{c}{$\sigma_B$}&
\multicolumn{1}{c}{V}&
\multicolumn{1}{c}{$\sigma_V$}&
\multicolumn{1}{c}{I}&
\multicolumn{1}{c}{$\sigma_I$}&
\multicolumn{1}{c}{Comments}}
\startdata
00&13:25:45.38&-42:58:47.44&------&3.52E+36&198&1670&24.40&0.11&----&----&----&----\nl
01&13:25:43.25&-42:58:37.26&29.9E+36&21.8E+36&313&1718&20.45&0.03&19.40&0.03&18.23&0.03\nl
02&13:25:38.62&-42:59:19.56&10.5E+36&1.01E+36&567&1510&19.83&0.04&18.90&0.03&17.84&0.03\nl
03&13:25:38.18&-42:58:15.26&52.3E+36&3.46E+36&----&----&----&----&----&----&----&----&sat GC\nl
04&13:25:35.52&-42:59:35.16&52.8E+36&48.4E+36&737&1432&20.48&0.03&19.41&0.02&18.25&0.02\nl
05&13:25:33.95&-42:58:59.53&48.2E+36&55.3E+36&----&----&----&----&----&----&----&----&sat GC\nl
06&13:25:33.38&-43:00:52.73&33.7E+36&49.4E+36&----&----&----&----&----&----&----&----&H$_{\alpha}$\nl
07&13:25:33.38&-42:59:13.02&5.29E+36&8.59E+36&----&----&----&----&----&----&----&----\nl
08&13:25:32.44&-42:58:49.98&23.9E+36&21.7E+36&904&1657&19.83&0.03&18.79&0.02&17.59&0.01\nl
09&13:25:31.61&-43:00:03.06&94.5E+36&80.0E+36&950&1294&18.39&0.05&----&----&----&----&sat GC\nl
10&13:25:30.57&-42:59:14.57&1.53E+36&240.E+36&----&----&----&----&----&----&----&----\nl
11&13:25:30.29&-42:59:34.64&1.96E+36&11.0E+36&1022&1436&19.71&0.05&18.91&0.04&17.93&0.03\nl
12&13:25:28.96&-42:59:30.83&4.72E+36&6.27E+36&----&----&----&----&----&----&----&----\nl
13&13:25:28.75&-42:59:48.23&113.E+36&95.7E+36&1106&1368&24.45&0.19&----&----&----&----\nl
14&13:25:26.96&-43:00:52.57&17.4E+36&14.6E+36&----&----&----&----&----&----&----&----\nl
15&13:25:26.71&-43:00:59.56&------&33.7E+36&1216&1014&22.73&0.04&23.13&0.35&----&----&H$_{\alpha}$\nl
16&13:25:26.43&-43:00:54.40&126.E+36&141.E+36&----&----&----&----&----&----&----&----\nl
17&13:25:25.76&-43:00:55.98&84.7E+36&166.E+36&1268&1030&21.70&0.10&20.83&0.07&19.37&0.11\nl
18&13:25:24.87&-43:00:02.05&3.94E+36&12.6E+36&----&----&----&----&----&----&----&----\nl
19&13:25:24.20&-42:59:59.41&60.5E+36&56.0E+36&1354&1312&23.48&0.12&22.36&0.09&21.13&0.07&H$_{\alpha}$\nl
20&13:25:23.96&-43:00:59.39&26.8E+36&36.7E+36&----&----&----&----&----&----&----&----\nl
21&13:25:23.66&-43:00:09.67&130.E+36&------&----&----&----&----&----&----&----&----&H$_{\alpha}$\nl
22&13:25:20.87&-43:00:56.96&6.73E+36&------&1635&1913&23.91&0.08&23.60&0.05&----&----\nl
23&13:25:20.86&-43:00:53.65&------&13.8E+36&1536&1042&22.70&0.12&21.99&0.11&20.90&0.11\nl
24&13:25:19.90&-43:00:53.28&------&8.35E+36&1590&1037&22.88&0.08&----&----&----&----\nl
25&13:25:19.05&-42:57:58.59&3.30E+36&5.01E+36&----&----&----&----&----&----&----&----\nl
26&13:25:42.13&-43:03:19.73&5.24E+36&2.94E+36&376&316&21.05&0.05&20.14&0.03&19.11&0.03\nl
27&13:25:40.84&-43:02:46.85&13.8E+36&20.3E+36&----&----&----&----&----&----&----&----\nl
28&13:25:40.82&-43:03:43.64&3.42E+36&------&----&----&----&----&----&----&----&----\nl
29&13:25:40.57&-43:01:15.05&44.8E+36&58.3E+36&----&----&----&----&----&----&----&----\nl
30&13:25:40.51&-43:02:51.30&4.35E+36&3.40E+36&464&458&20.35&0.05&19.00&0.04&17.58&0.03&H$_{\alpha}$\nl
31&13:25:38.33&-43:02:05.81&208.E+36&197.E+36&584&684&23.99&0.09&----&----&----&----\nl
32&13:25:37.54&-43:01:42.41&4.10E+36&2.27E+36&----&----&----&----&----&----&----&----&H$_{\alpha}$\nl
33&13:25:37.46&-43:01:31.43&6.32E+36&7.08E+36&630&854&19.12&0.03&----&----&----&----&sat GC\nl
34&13:25:36.64&-43:00:57.53&9.59E+36&33.5E+36&674&1024&23.60&0.09&----&----&----&----\nl
35&13:25:35.23&-43:02:33.56&10.3E+36&6.88E+36&----&----&----&----&----&----&----&----\nl
36&13:25:33.68&-43:03:13.32&17.9E+36&17.2E+36&836&348&23.84&0.10&22.97&0.04&21.18&0.04\nl
37&13:25:33.09&-43:01:08.21&24.6E+36&23.9E+36&----&----&----&----&----&----&----&----\nl
38&13:25:32.48&-43:01:34.14&69.2E+36&68.0E+36&----&----&----&----&----&----&----&----\nl
39&13:25:32.35&-43:01:27.19&------&4.58E+36&910&874&22.87&0.11&22.20&0.10&----&----&H$_{\alpha}$\nl
40&13:25:32.02&-43:02:31.57&53.1E+36&54.7E+36&928&556&20.32&0.06&19.09&0.04&17.76&0.04&H$_{\alpha}$\nl
41&13:25:31.97&-43:03:02.61&4.89E+36&3.35E+36&----&----&----&----&----&----&----&----\nl
42&13:25:31.63&-43:01:18.90&6.78E+36&------&948&917&22.31&0.07&21.48&0.08&----&----&H$_{\alpha}$\nl
43&13:25:31.34&-43:02:03.98&23.3E+36&1.91E+36&----&----&----&----&----&----&----&----\nl
44&13:25:29.45&-43:01:08.36&47.9E+36&32.4E+36&1068&968&19.65&0.07&18.70&0.07&17.39&0.04\nl
45&13:25:29.23&-43:01:14.78&------&12.9E+36&1080&938&23.77&0.14&----&----&----&----&H$_{\alpha}$\nl
46&13:25:28.45&-43:03:15.47&25.8E+36&25.7E+36&----&----&----&----&----&----&----&----\nl
47&13:25:28.32&-43:04:16.56&5.47E+36&6.91E+36&----&----&----&----&----&----&----&----\nl
48&13:25:28.25&-43:02:53.52&31.1E+36&33.2E+36&1133&447&21.86&0.03&21.36&0.01&21.03&0.02\nl
49&13:25:28.06&-43:01:18.51&------&14.8E+36&1143&919&23.29&0.10&22.06&0.08&----&----&H$_{\alpha}$\nl
50&13:25:28.04&-43:04:02.81&19.1E+36&19.8E+36&1145&102&20.48&0.04&19.39&0.03&18.12&0.03\nl
51&13:25:27.71&-43:02:17.80&7.92E+36&13.8E+36&----&----&----&----&----&----&----&----&sat fgs\nl
52&13:25:27.45&-43:02:14.10&165.E+36&83.7E+36&1177&643&21.23&0.13&----&----&----&----\nl
53&13:25:27.49&-43:01:28.43&47.9E+36&61.5E+36&----&----&----&----&----&----&----&----\nl
54&13:25:27.10&-43:01:59.34&55.7E+36&69.1E+36&----&----&----&----&----&----&----&----&H$_{\alpha}$\nl
55&13:25:25.59&-43:02:09.68&7.33E+36&14.4E+36&1277&665&22.70&0.07&21.71&0.05&20.48&0.03\nl
56&13:25:25.26&-43:02:23.01&17.4E+36&21.8E+36&----&----&----&----&----&----&----&----\nl
57&13:25:25.15&-43:01:26.81&30.1E+36&28.5E+36&1301&878&20.82&0.03&19.44&0.02&18.95&0.01\nl
58&13:25:24.88&-43:04:25.24&21.0E+36&19.7E+36&----&----&----&----&----&----&----&----\nl
59&13:25:24.76&-43:01:24.77&23.2E+36&18.8E+36&----&----&----&----&----&----&----&----\nl
60&13:25:24.51&-43:01:41.88&12.1E+36&11.6E+36&1336&803&23.32&0.15&----&----&----&----&H$_{\alpha}$\nl
61&13:25:24.39&-43:01:09.96&------&10.3E+36&----&----&----&----&----&----&----&----\nl
62&13:25:23.62&-43:03:26.09&37.1E+36&40.8E+36&----&----&----&----&----&----&----&----\nl
63&13:25:23.57&-43:02:20.72&29.1E+36&34.3E+36&1388&610&23.16&0.05&22.36&0.04&20.74&0.03\nl
64&13:25:23.54&-43:01:38.45&49.7E+36&58.8E+36&1390&820&22.77&0.11&----&----&----&----\nl
65&13:25:23.10&-43:01:34.39&24.7E+36&22.9E+36&1414&840&24.03&0.14&----&----&----&----\nl
66&13:25:23.06&-43:01:45.81&27.3E+36&22.2E+36&----&----&----&----&----&----&----&----\nl
67&13:25:22.88&-43:01:24.95&174.E+36&92.7E+36&1426&888&22.25&0.15&----&----&----&----\nl
68&13:25:21.75&-43:01:54.15&7.77E+36&8.25E+36&----&----&----&----&----&----&----&----\nl
69&13:25:21.56&-43:02:13.73&14.5E+36&16.0E+36&----&----&----&----&----&----&----&----\nl
70&13:25:21.27&-43:01:58.80&21.9E+36&13.5E+36&----&----&----&----&----&----&----&----\nl
71&13:25:20.09&-43:03:10.10&------&30.9E+36&1578&364&21.31&0.03&20.19&0.02&18.88&0.01\nl
72&13:25:19.87&-43:03:17.14&1120E+36&2.10E+36&----&----&----&----&----&----&----&----\nl
73&13:25:19.19&-43:01:57.35&18.4E+36&19.3E+36&----&----&----&----&----&----&----&----\nl
74&13:25:18.50&-43:01:16.34&41.7E+36&43.0E+36&1664&930&19.47&0.03&18.20&0.02&----&----\nl
75&13:25:17.87&-43:02:05.09&42.8E+36&29.3E+36&----&----&----&----&----&----&----&----\nl
76&13:25:16.40&-43:02:55.25&37.1E+36&16.2E+36&----&----&----&----&----&----&----&----\nl
77&13:25:15.79&-43:01:58.10&5.54E+36&8.74E+36&1812&722&22.03&0.02&21.68&0.02&20.69&0.02\nl
78&13:25:15.82&-42:57:39.66&15.6E+36&10.9E+36&----&----&----&----&----&----&----&----\nl
79&13:25:14.04&-43:01:21.22&5.28E+36&7.01E+36&----&----&----&----&----&----&----&----\nl
80&13:25:14.03&-43:02:42.90&7.46E+36&1.87E+36&1903&496&21.20&0.02&20.40&0.03&19.38&0.02\nl
81&13:25:14.03&-42:59:50.33&7.25E+36&5.46E+36&----&----&----&----&----&----&----&----\nl
82&13:25:12.89&-43:01:14.69&89.4E+36&75.9E+36&1971&938&21.17&0.03&20.00&0.02&18.66&0.01\nl
83&13:25:12.36&-43:00:49.55&5.33E+36&6.97E+36&----&----&----&----&----&----&----&----\nl
84&13:25:12.04&-43:02:20.47&4.76E+36&11.6E+36&----&----&----&----&----&----&----&----\nl
85&13:25:12.00&-43:00:10.79&40.7E+36&38.9E+36&2020&1256&24.08&0.09&22.51&0.04&20.79&0.03&fgs\nl
86&13:25:11.99&-43:00:44.63&44.8E+36&36.3E+36&----&----&----&----&----&----&----&----\nl
87&13:25:11.52&-43:02:27.09&15.4E+36&17.6E+36&----&----&----&----&----&----&----&----\nl
88&13:25:48.53&-43:03:23.19&9.30E+36&4.95E+36&----&----&----&----&----&----&----&----\nl
89&13:25:47.24&-42:58:25.13&18.8E+36&17.5E+36&----&----&----&----&----&----&----&----\nl
90&13:25:47.19&-43:02:43.45&8.42E+36&12.3E+36&100&497&23.90&0.07&22.69&0.03&21.69&0.03\nl
91&13:25:46.57&-43:00:35.99&------&8.72E+36&----&----&----&----&----&----&----&----\nl
92&13:25:46.38&-43:03:10.61&22.2E+36&23.2E+36&144&361&24.24&0.10&22.58&0.04&20.70&0.03&fgs\nl
93&13:25:45.70&-43:01:15.90&2.68E+36&4.44E+36&181&932&19.84&0.05&18.97&0.04&18.02&0.03\nl
94&13:25:45.45&-42:58:15.59&12.3E+36&3.01E+36&----&----&----&----&----&----&----&----\nl
101&13:25:32.88&-43:00:31.00&73.0E+36&73.0E+36&883&1164&24.15&0.17&23.36&0.13&22.18&0.10\nl
102&13:25:34.13&-43:00:10.20&20.6E+36&20.6E+36&815&1268&24.42&0.50&----&----&----&----\nl
\enddata
\tablerefs{ \\
$L_X$316 and $L_X$962 fluxes in ergs/sec from Kraft et al. 2001. \\
$B$, $V$, and $I$-band data from ESO VLT.\\
sat: Saturated.\\
fgs: Foreground star.\\
GC: Globular cluster.\\
H$\alpha$: Detected in H$\alpha$.\\
}
\end {planotable}

\begin{planotable}{llllllll}
\small
\footnotesize
\tablewidth{0pt}
\scriptsize
\tablecaption{Additional Optical ID of NGC5128 X-Ray Point Sources}
\tablehead{
\multicolumn{1}{c}{MRF}&
\multicolumn{1}{c}{RA}&
\multicolumn{1}{c}{DEC}&
\multicolumn{1}{c}{$L_X$316}&
\multicolumn{1}{c}{$L_X$962}&
\multicolumn{1}{c}{Comments}}
\startdata
216&13:25:07.65&-42:56:30.23&29.0E+36&------- &GC? \nl
215&13:25:09.20&-42:58:59.64&62.6E+36&51.1E+36&GC? \nl
213&13:25:10.09&-42:56:07.89&15.7E+36&------- &FC \nl
212&13:25:11.93&-42:57:12.82&9.98E+36&10.4E+36&FC \nl
210&13:25:18.80&-42:57:08.08&11.4E+36&5.47E+36& \nl
209&13:25:20.19&-42:56:15.26&10.1E+36&6.96E+36&FC \nl
208&13:25:22.36&-42:57:17.09&85.8E+36&71.2E+36&GC? \nl
207&13:25:23.49&-42:56:51.54&21.8E+36&21.1E+36& \nl
206&13:25:32.50&-42:56:23.90&3.98E+36&------- & \nl
205&13:25:32.80&-42:56:23.83&5.98E+36&10.4E+36&FC \nl
204&13:25:34.43&-42:55:49.61&9.09E+36&6.80E+36& \nl
203&13:25:38.43&-42:56:30.48&8.92E+36&9.04E+36& \nl
202&13:25:39.10&-42:56:53.52&55.3E+36&29.8E+36&FC \nl
201&13:25:39.39&-42:55:46.34&4.01E+36&1.51E+36& \nl
\enddata
\tablerefs{ \\
$L_X$316 and $L_X$962 fluxes in ergs/sec from Kraft et al. 2001. \\
FC: Faint optical counterpart detected in VLT images.\\
GC?: Possible globular cluster.\\
}
\end {planotable}

\begin{planotable}{llllllllllllll}
\small
\footnotesize
\tablewidth{0pt}
\scriptsize
\tablecaption{NGC5128 Globular Cluster X-Ray Sources}
\tablehead{
\multicolumn{1}{c}{MRF}&
\multicolumn{1}{c}{RA}&
\multicolumn{1}{c}{DEC}&
\multicolumn{1}{c}{$L_X$316}&
\multicolumn{1}{c}{$L_X$962}&
\multicolumn{1}{c}{B}&
\multicolumn{1}{c}{V}&
\multicolumn{1}{c}{I}&
\multicolumn{1}{c}{Spectrum}&
\multicolumn{1}{c}{Reference}}
\startdata
 01&13:25:43.25&-42:58:37.26&29.9E+36&21.8E+36&20.45&19.40&18.23&&\nl
 02&13:25:38.62&-42:59:19.56&10.5E+36&1.01E+36&19.83&18.90&17.84&yes&Harris+ 1992\nl
 03&13:25:38.18&-42:58:15.26&52.3E+36&3.46E+36&----&----&----&yes&sat\nl
 04&13:25:35.52&-42:59:35.16&52.8E+36&48.4E+36&20.48&19.41&18.25&&\nl
 05&13:25:33.95&-42:58:59.53&48.2E+36&55.3E+36&18.21&17.73&----&yes&Harris+ 1992\nl
 08&13:25:32.44&-42:58:49.98&23.9E+36&21.7E+36&19.83&18.79&17.59&yes&Minniti+ 1996\nl
 09&13:25:31.61&-43:00:03.06&94.5E+36&80.0E+36&18.39&16.83&----&&Minniti+ 1996\nl
 11&13:25:30.29&-42:59:34.64&1.96E+36&11.0E+36&19.71&18.91&17.93&yes&Minniti+ 1996\nl
 17&13:25:25.76&-43:00:55.98&84.7E+36&166.E+36&21.17&20.83&19.37&&\nl
 23&13:25:20.86&-43:00:53.65&------  &13.8E+36&22.70&21.89&20.90&&\nl
 24&13:25:19.90&-43:00:53.28&------  &8.35E+36&----&----&----&&sat\nl
 26&13:25:42.13&-43:03:19.73&5.24E+36&2.94E+36&21.05&20.14&19.11&&\nl
 30&13:25:40.51&-43:02:51.30&4.35E+36&3.40E+36&20.35&19.00&17.58&&\nl
 33&13:25:37.46&-43:01:31.43&6.32E+36&7.08E+36&19.12&----&----&yes&sat\nl
 40&13:25:32.02&-43:02:31.57&53.1E+36&54.7E+36&20.32&19.09&17.76&&\nl
 44&13:25:29.45&-43:01:08.36&47.9E+36&32.4E+36&19.65&18.70&17.39&&\nl
 49&13:25:28.06&-43:01:18.51&  ------&14.8E+36&23.29&22.06&----&&\nl
 50&13:25:28.04&-43:04:02.81&19.1E+36&19.8E+36&20.48&19.39&18.12&&\nl
53&13:25:27.49&-43:01:28.43&47.9E+36&61.5E+36&17.78&17.57&----&yes&Harris+ 1992 \nl
 55&13:25:25.59&-43:02:09.68&7.33E+36&14.4E+36&22.70&21.71&20.48&&\nl
 57&13:25:25.15&-43:01:26.81&30.1E+36&28.5E+36&20.82&19.44&18.05&&\nl
 71&13:25:20.09&-43:03:10.10&  ------&30.9E+36&21.31&20.19&18.88&&\nl
 74&13:25:18.50&-43:01:16.34&41.7E+36&43.0E+36&19.47&18.20&----&&\nl
 80&13:25:14.03&-43:02:42.90&7.46E+36&1.87E+36&21.20&20.40&19.38&&\nl
 82&13:25:12.89&-43:01:14.69&89.4E+36&75.9E+36&21.17&20.00&18.66&&\nl
 93&13:25:45.70&-43:01:15.90&2.68E+36&4.44E+36&19.84&18.97&18.02&&\nl
216&13:25:07.65&-42:56:30.23&29.0E+36&------- &----&----&----&&GC? \nl
215&13:25:09.20&-42:58:59.64&62.6E+36&51.1E+36&----&----&----&&GC? \nl
208&13:25:22.36&-42:57:17.09&85.8E+36&71.2E+36&----&----&----&&GC? \nl
f1.16&13:26:10.58&-42:53:42.78&21.3E+36&  ------&----&18.47&----&yes&Rejkuba 2001 \nl
f2.81&13:25:05.75&-43:10:31.51&  ------&14.1E+36&----&18.01&----&yes&Rejkuba 2001 \nl
G176&13:25:03.04&-42:56:25.09&17.0E+36&  ------&19.97&18.87&----&yes&Harris+ 1992 \nl
G284&13:25:46.59&-42:57:02.69&38.6E+36&42.7E+36&20.96&19.70&----&yes&Harris+ 1992 \nl
\enddata
\tablerefs{ \\
$L_X$316 and $L_X$962 in ergs/sec. \\
$B$, $V$, and $I$-band data from ESO VLT.\\
* Source 53 is not matched in our frames, GC association is from Kraft et al. 2001.\\
sat = saturated\\
GC? = no colors available
}
\end {planotable}

\end{document}